# Walking and running with non-specific chronic low back pain: what about the lumbar lordosis angle?

Edwige SIMONET[1], Balz WINTELER[1,2], Jana FRANGI[1], Magdalena SUTER[1,3], Michael L. MEIER[3], Patric EICHELBERGER[1], Heiner BAUR[1], Stefan SCHMID[1,*]

[1]Bern University of Applied Sciences, Department of Health Professions, Division of Physiotherapy, Spinal Movement Biomechanics Group, Bern, Switzerland
[2]Bern University Hospital, Inselspital, Department of Physiotherapy, Bern, Switzerland
[3]Balgrist University Hospital, Department of Chiropractic Medicine, Integrative Spinal Research, Zurich, Switzerland

**Corresponding author:**
[*]Stefan Schmid, PT, PhD, Bern University of Applied Sciences, Department of Health Professions, Murtenstrasse 10, 3008 Bern, Switzerland, +41 79 936 74 79, stefanschmid79@gmail.com

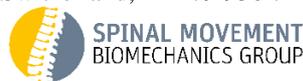

**ABSTRACT**

Non-specific chronic low back pain (NSCLBP) is a major health problem, affecting about one fifth of the population worldwide. To avoid further pain or injury, patients with NSCLBP seem to adopt a stiffer movement pattern during everyday living activities. However, it remains unknown how NSCLBP affects the lumbar lordosis angle (LLA) during repetitive activities such as walking or running. This pilot study therefore aimed at exploring possible NSCLBP-related alterations in LLAs during walking and running by focusing on discrete parameters as well as continuous data.

Thirteen patients with NSCLBP and 20 healthy pain-free controls were enrolled and underwent a full-body movement analysis involving various everyday living activities such as standing, walking and running. LLAs were derived from markers placed on the spinous processes of the vertebrae L1-L5 and S1. Possible group differences in discrete (average and range of motion (ROM)) and continuous LLAs were analyzed descriptively using mean differences with confidence intervals ranging from 95% to 75%.

Patients with NSCLBP indicated reduced average LLAs during standing, walking and running and a tendency for lower LLA-ROM during walking. Analyses of continuous data indicated the largest group differences occurring around 25% and 70% of the walking and 25% and 75% of the running cycle. Furthermore, patients indicated a reversed movement pattern during running, with increasing instead of a decreasing LLAs after foot strike.

This study provides preliminary evidence that NSCLBP might affect LLAs during walking and running. These results can be used as a basis for future large-scale investigations involving hypothesis testing.

# 1. INTRODUCTION

Low back pain (LBP) is a major global health problem, with up to 84% of the population having at least one episode of LBP during their lifetime and 23% developing a chronic form (Balague et al., 2012). In about 9 out of 10 patients with chronic LBP, the pathoanatomical cause cannot be determined and the term non-specific chronic LBP (NSCLBP) is used (Machado et al., 2009; Maher et al., 2017). Clinical guidelines recommend patient education, medication as well as exercise therapy and spinal manipulation to treat NSCLBP (Balague et al., 2012; Chou et al., 2007). However, the effect sizes for most treatment options compared to placebo were relatively small, indicating a need for more effective approaches (Balague et al., 2012; Keller et al., 2007; Machado et al., 2009). In order to develop better treatment and prevention strategies, a profound understanding of the underlying mechanisms of NSCLBP is of high importance.

To avoid further pain or injury, people move differently. Adaptions in trunk movement and their potential impact on LBP are gaining more and more attention as evolving evidence indicates that certain spinal movement strategies during daily activities might predispose individuals to persistent back problems in the long term through disruption of the load sharing of spinal tissues and muscle fatigue, promoting pro-nociceptive mechanisms and injury (Hodges and Smeets, 2015; van Dieen et al., 2017). As such, throughout the past decade, several research groups have attempted to better understand lumbar spine function in NSCLBP, especially by studying everyday living activities such as walking, chair rising, object lifting, stepping up and running (Christe et al., 2017; Christe et al., 2016; Christe et al., 2019; Gombatto et al., 2015; Gombatto et al., 2017; Hemming et al., 2018; Hernandez et al., 2017; Muller et al., 2015; Papi et al., 2019; Pelegrinelli et al., 2020; van den Hoorn et al., 2012). However, these studies were based on differently defined rigid trunk segments and none of them considered the lumbar lordosis angle, which is a parameter that complements the findings of the above mentioned studies by providing additional information that might not be detected using rigid segments (Schmid et al., 2016). Moreover, the lumbar lordosis angle is a clinical parameter that is familiar to every clinician and might hence be fairly easily addressed in prevention or treatment protocols. Finally, and very importantly, they reduced the kinematic data to discrete parameters such as range of motion (RoM) and did not consider the continuous data over time, which might have led to the missing of valuable information.

Among several types of physical activity, walking and running are of the least expensive and most broadly accessible forms of physical activity and were shown to be associated with various health benefits (Hamer and Chida, 2008; Pedisic et al., 2019). Even though patients with LBP seem not to be less active than healthy individuals (Griffin et al., 2012), it is still not fully understood how LBP affects lumbar spine movement or vice versa during walking and running.

For these reasons, this pilot study aimed at exploring possible differences in lumbar lordosis angles during walking and running between patients with NSCLBP and healthy pain-free controls by focusing on discrete parameters as well as continuous data.

# 2. METHODS

## 2.1. Study population

Thirteen patients with NSCLBP and 20 healthy controls were included in this pilot study (Table 1). The recommendations provided by Whitehead et al. (2016) suggested a minimum sample size of n=12 per group when assuming a target effect size of 0.5, a power of 90% and a Type I error rate of 5% for a future larger trial. Participants were included if they were aged 18-60 years. Patients had to suffer from NSCLBP for at least three months and the healthy controls had to be free of pain with no prior history of NSCLBP. Exclusion criteria for both groups were a history of specific LBP or nerve root pain, any disorders affecting the gait or running patterns, severe psychological disorders, anamnestically known pregnancy and breastfeeding less than 6 months postnatal as well as a body mass index (BMI) > $30kg/m^2$. Additional exclusion criteria for the patients with NSCLBP were intensive sporting activities of more than twice a week, a current back pain level of > 5/10 points on the visual analog scale (VAS) (Jensen et al., 2003) and an Oswestry Disability Index (ODI) of > 40% (Davidson and Keating, 2002). The study protocol was approved by the local ethics committee and all participants provided written informed consent.

**Table 1:** Demographics of the healthy pain-free controls and the patients with non-specific chronic low back pain (NSCLBP). Presented are mean with one standard deviation (SD) and range (in brackets). Group differences were investigated using independent samples T-tests with an alpha-level set to 0.05.

| Characteristics | Controls (n=20) | NSCLBP patients (n=13) | Group diff. (p) |
|---|---|---|---|
| Age [years] | 31.4 SD 9.2 (20.0-53.0) | 38.0 SD 11.6 (23.0-54.0) | 0.095 |
| Height [cm] | 173.9 SD 9.6 (157.0-192.0) | 174.6 SD 7.3 (161.0-188.3) | 0.794 |
| Mass [kg] | 68.9 SD 12.9 (45.5-91.7) | 67.0 SD 12.0 (48.7-91.0) | 0.669 |
| Gender [m/f] | 9/11 | 8/5 | N/A |
| VAS[1] [0-10] | N/A | 2.3 SD 1.1 (1-4) | N/A |
| ODI[2] [%] | N/A | 19.5 SD 9.2 (12-34) | N/A |

[1]Visual analog scale (VAS)
[2]Oswestry Disability Index (ODI)

### 2.2. Experimental procedures

All participants were invited for a visit at the movement analysis laboratory. Patients were first asked to rate their current back pain level using a VAS (0=no pain, 10=worst pain imaginable) and to complete the German version of the ODI (0%=no disability, 100%= bed-bound or exaggerating their symptoms) to quantify their perceived level of LBP-related disability (Jensen et al., 2003; Mannion et al., 2006). Subsequently, all participants were equipped with 58 retro-reflective markers by two trained physiotherapists based on a previously described configuration (Schmid et al., 2017). To allow for a detailed quantification of spinal motion, the configuration includes markers on the spinous processes of C7, T3, T5, T7, T9, T11 as well as L1-5 and S1 (sacrum). Participants were then asked to perform a variety of everyday living activities including upright standing for 10 seconds as well as walking and running on a 10-meter level walkway at a normal self-selected speed. Marker positions were tracked using a three-dimensional optical motion capture system with 10 infrared cameras (VICON, Oxford, UK; sampling rate: 200Hz). The standing activity was performed once, whereas the walking and running activities were repeated until five valid trials were recorded.

### 2.3. Data reduction

Data pre-processing was conducted using the software package Vicon Nexus 2.6 (Vicon UK, Oxford, UK), which included the reconstruction and filtering of the marker trajectories as well as event setting to identify the respective walking and running cycles. Lumbar lordosis angles were then calculated using a custom-built MATLAB-routine (R2019a, MathWorks Inc.,

Natick, MA, USA) as described previously (Schmid et al., 2015). In brief, we applied a combination of a quadratic polynomial and a circle fit function to the sagittal plane trajectories of the markers from L1 to S1 and used the central angle theorem to derive the lumbar lordosis angle (expressed in degrees). Outcome parameters were defined as average lumbar lordosis angles during upright standing, walking and running, RoM (absolute difference between maximum and minimum angle) of the lumbar lordosis angles during walking and running as well as continuous lumbar lordosis angles during walking and running (time-normalized to 101 data points).

## 2.4. Statistical analysis

All calculations were carried out using MATLAB (R2019a, MathWorks Inc., Natick, MA, USA). To explore possible differences in lumbar lordosis angles (discrete parameters and continuous data) between patients with NSCLBP and healthy pain-free controls, we calculated means and standard deviations as well as mean differences with 95%, 90%, 85%, 80% and 75% confidence intervals (CI) as suggested by Lee et al. (2014). Since our goal was not to test hypotheses but to provide initial data for future larger trials, we did not perform any calculations of probability (e.g. independent samples T-test).

## 3. RESULTS

### 3.1. Discrete parameters

In patients with NSCLBP, average lumbar lordosis angles during upright standing, walking and running appeared to be reduced by about 5-10 degrees with confidence levels of 75%, 90% and 75%, respectively (Figure 1). The RoM of the lumbar lordosis angle indicated a tendency for lower values during walking (about 2.5 degrees with a confidence level of 90%) but not during running.

### 3.2. Continuous data

Mean differences of the continuous data confirmed the generally lower lumbar lordosis angles in patients with NSCLBP, with the largest differences occurring around 25% and 70% of the walking and running cycles (confidence level at 85%) (Figure 2). In addition, the movement pattern of the lumbar lordosis angles during running appeared to be reversed in NSCLBP patients, with an increasing instead of a decreasing angle after foot strike.

## 4. DISCUSSION

This pilot study aimed at exploring possible alterations in lumbar lordosis angles during walking and running in patients with NSCLBP compared to healthy pain-free controls. The results indicated that patients with NSCLBP possibly adopt a generally less lordotic lumbar posture as well as a reduced motion during walking and a reversed movement pattern during running.
The attenuated lumbar lordosis angles in patients with NSCLBP are in line with the findings of a recent meta-analysis of 796 patients with LBP and 927 healthy controls (Chun et al., 2017). Since standing with more lumbar lordosis was shown to be related to a higher risk for developing or maintaining LBP (Sorensen et al., 2015), the observed lordosis attenuation in NSCLBP patients might be interpreted as a protective or compensatory mechanisms.

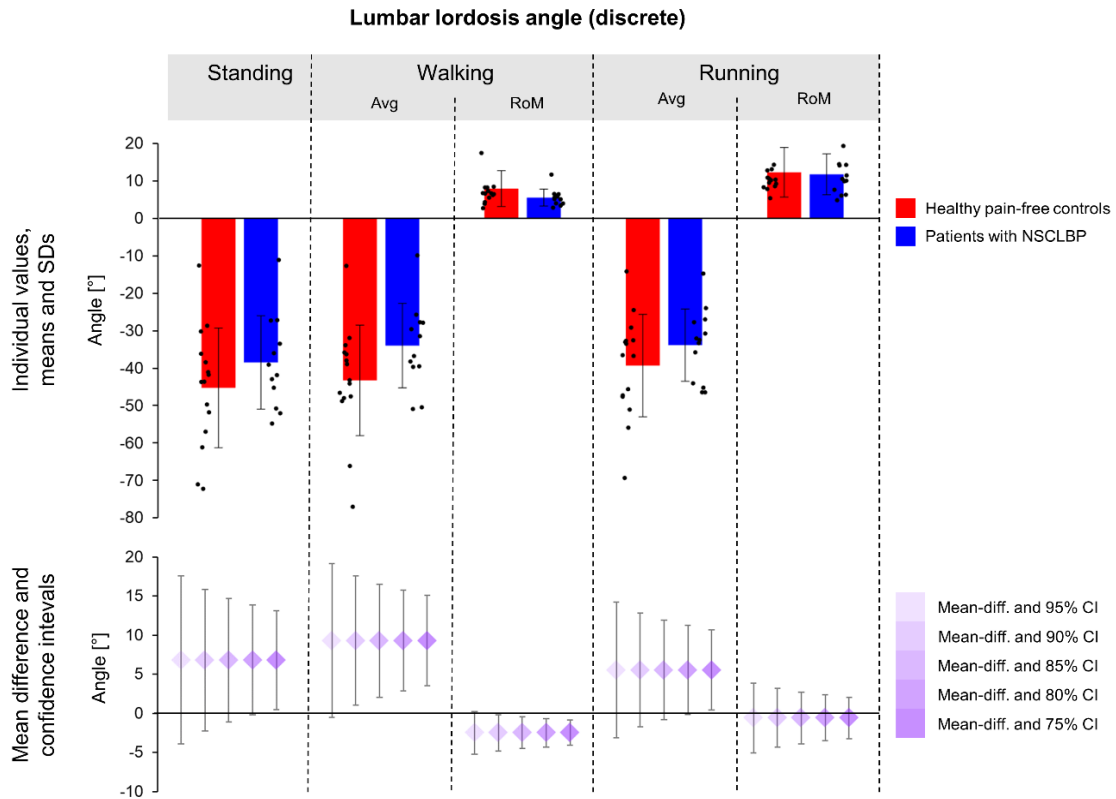

**Figure 1:** Individual values, means and standard deviations (SD) as well as mean differences with confidence intervals ranging from 95% to 75% for the average (Avg) and range of motion (RoM) lumbar lordosis angles in healthy pain-free controls and patients with non-specific chronic low back pain (NSCLBP) during upright standing, walking and running.

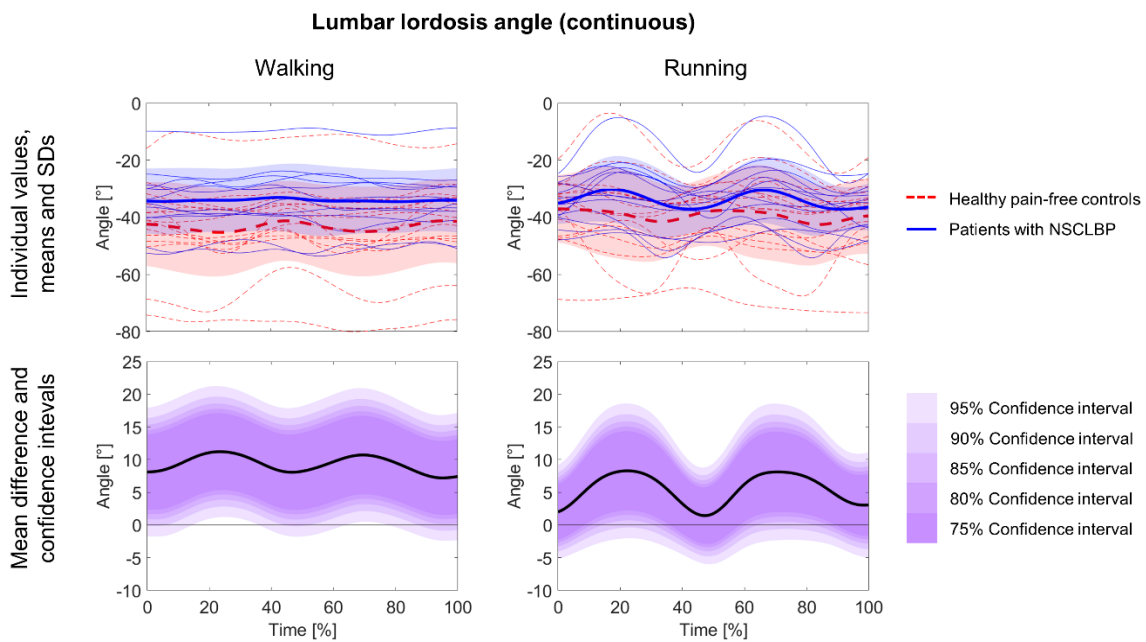

**Figure 2:** Individual values, means and standard deviations (SD) as well as mean differences with confidence intervals ranging from 95% to 75% for the continuous lumbar lordosis angles over one full cycle in healthy pain-free controls and patients with non-specific chronic low back pain (NSCLBP) during walking and running.

The reduced sagittal lumbar motion during walking agrees with the assumption that patients with LBP adopt protective movement strategies by increasing trunk stiffness (van den Hoorn et al., 2012). In fact, it was reported that individuals with chronic LBP increase their abdominal muscle activity particularly during the single stance phases of walking (Pakzad et al., 2016), which coincides with the phases where the patients with NSCLBP in our study seemed to avoid an increase in lumbar lordosis. The reversed movement pattern observed in our patients during running can probably be interpreted as an overcorrection of the trunk stiffening strategy and should be further investigated. Especially during an activity with relatively high impacts such as running, these biomechanical changes might be of high clinical importance.

It is important to mention though that about one fourth (23%) of the patients presented a normal lumbar lordosis movement pattern, i.e. a movement pattern that was comparable to the majority of healthy controls. On the other side, one fifth (20%) of healthy controls presented a movement pattern that resembled the majority of patients with NSCLBP, i.e. a reversed movement pattern, indicating different trunk motor control strategies among groups. A possible explanation for this might be found when looking at psychomotor interactions, e.g. the interplay between fear-avoidance beliefs and spinal motion in individual participants. Studies investigating spinal motion during object lifting in patients with NSCLBP as well as in healthy pain-free adults showed that individuals who believed that lifting an object with a round back is dangerous had altered lumbar spine kinematics (Matheve et al., 2019; Schmid et al., 2019). Unfortunately, literature lacks comparable evidence on running and since the current study did not include any assessments of fear-avoidance beliefs, this explanation attempt remains speculative.

A limiting factor of this study was that in a few cases (5 healthy controls during standing, walking and running; 1 patient during standing and walking), our 10-camera motion capture system was not able to fully identify all the lumbar spine markers and hence, no lumbar lordosis angles could be calculated in these cases. For future studies, it is therefore advised to use more cameras or cameras with a higher resolution in order to avoid data loss due to technical limitations. Furthermore, even though the average ages of the two groups did not show statistically significant differences, patients appeared to be slightly older than the healthy controls, which could have influenced the outcomes. However, it is not expected that this age difference would explain a reversed movement pattern during running.

In summary, this study provides evidence that NSCLBP might affect the lumbar lordosis angle during walking and running. These results can be used as a basis for future large-scale investigations involving hypothesis testing. To further explore the reversed movement pattern during running, it does not make sense to use RoM as an outcome parameter but rather mean differences of the lumbar lordosis angles at about 25% and 75% of the running cycle.

## 5. CONFLICT OF INTEREST STATEMENT

The authors declare no conflict of interest.

## 6. ACKNOWLEDGMENTS

The authors thank the Swiss Physiotherapy Association (physioswiss) for financial assistance.